\documentclass[twoside,12pt]{article}

\usepackage{braket}
\usepackage[T1]{fontenc}
\usepackage{url}
\usepackage{bm}
\usepackage{subfig}
\usepackage{float}
\usepackage[pdftex]{graphicx}
\graphicspath{{figures_proc/}}

\def\Journal#1#2#3#4{{#1} {#2} (#4) #3 }

\def\NPA{{\em Nucl. Phys.} A}

\def\PLB{{\em Phys. Lett.} B}

\def\PRL{\em Phys. Rev. Lett.}

\def\PREP{\em Phys. Rep.}

\def\PRD{{\em Phys. Rev.} D}
\def\PRC{{\em Phys. Rev.} C}

\newcommand{\Tr}{\mbox{\rm Tr}}
\newcommand{\ReC}{\mbox{\rm Re}}

\newcommand{\be}{\begin{equation}}
\newcommand{\ee}{\end{equation}}
\newcommand{\bea}{\begin{eqnarray}}
\newcommand{\eea}{\end{eqnarray}}
\newcommand{\non}{\nonumber}
\newcommand{\bei}{\begin{itemize}}
\newcommand{\eei}{\end{itemize}}
\newcommand{\mbf}{\mathbf}

\begin{document}

\title{ \vspace{1cm} Colour flux-tubes in static Pentaquark and Tetraquark systems}
\author{Pedro Bicudo,$^1$ Nuno Cardoso,$^1$ Marco Cardoso$^1$\\
\\
$^1$CFTP, Instituto Superior T\'ecnico, Universidade T\'ecnica de Lisboa}

\maketitle
\begin{abstract} The colour fields created by the static tetraquark and pentaquark systems are computed in quenched SU(3) lattice QCD, with gauge invariant lattice
operators, in a $24^3 \times 48$ lattice at $\beta=6.2$.
We generate our quenched configurations with GPUs, and detail the respective benchmanrks in different SU(N) groups.
While at smaller distances the coulomb potential is expected to dominate, at larger distances it is
expected that fundamental flux tubes, similar to the flux-tube between a quark and an antiquark, emerge and confine the quarks. In order to minimize the
potential the fundamental flux tubes should connect at 120o angles. We compute the square of the colour fields utilizing plaquettes, and locate the static
sources with generalized Wilson loops and with APE smearing. The tetraquark system is well described by a double-Y-shaped flux-tube, with two Steiner
points, but when quark-antiquark pairs are close enough the two junctions collapse and we have an X-shaped flux-tube, with one Steiner point. The
pentaquark system is well described by a three-Y-shaped flux-tube where the three flux the junctions are Steiner points.
\end{abstract}
\section{Motivation}

Multiquark exotic hadrons like the tetraquark and the pentaquark, different from the the ordinary mesons and baryons, have been studied and searched for many years.
The tetraquark was initially proposed by Jaffe \cite{Jaffe:1976ig} as a bound state formed by two quarks and two antiquarks.
Presently several observed resonances are tetraquark candidates.
The most recent tetraquark candidates have been reported by the Belle Collaboration in May, the charged bottomonium
$Z_b^+(10610)$ and $Z_b^+(10650)$
\cite{Collaboration:2011gja}.
However a better understanding of tetraquarks is necessary to confirm or disprove the  X, Y and possibly also light resonances candidates as tetraquark states.

\begin{figure}[t!]
\begin{centering}
    \subfloat[\label{fig:tq1}]{
\begin{centering}
{
    \includegraphics[width=5.5cm]{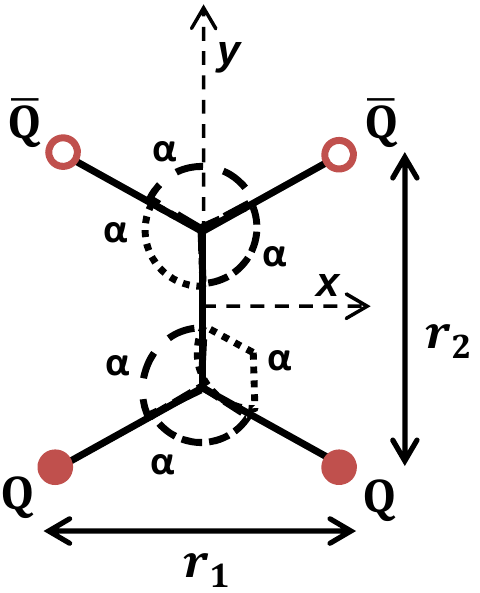}}
\par\end{centering}}
\subfloat[\label{fig:TQ_EB_ape_hyp_r1_8_r2_14_Act_3D_Sim}]{
\begin{centering}
    \includegraphics[width=9.cm]{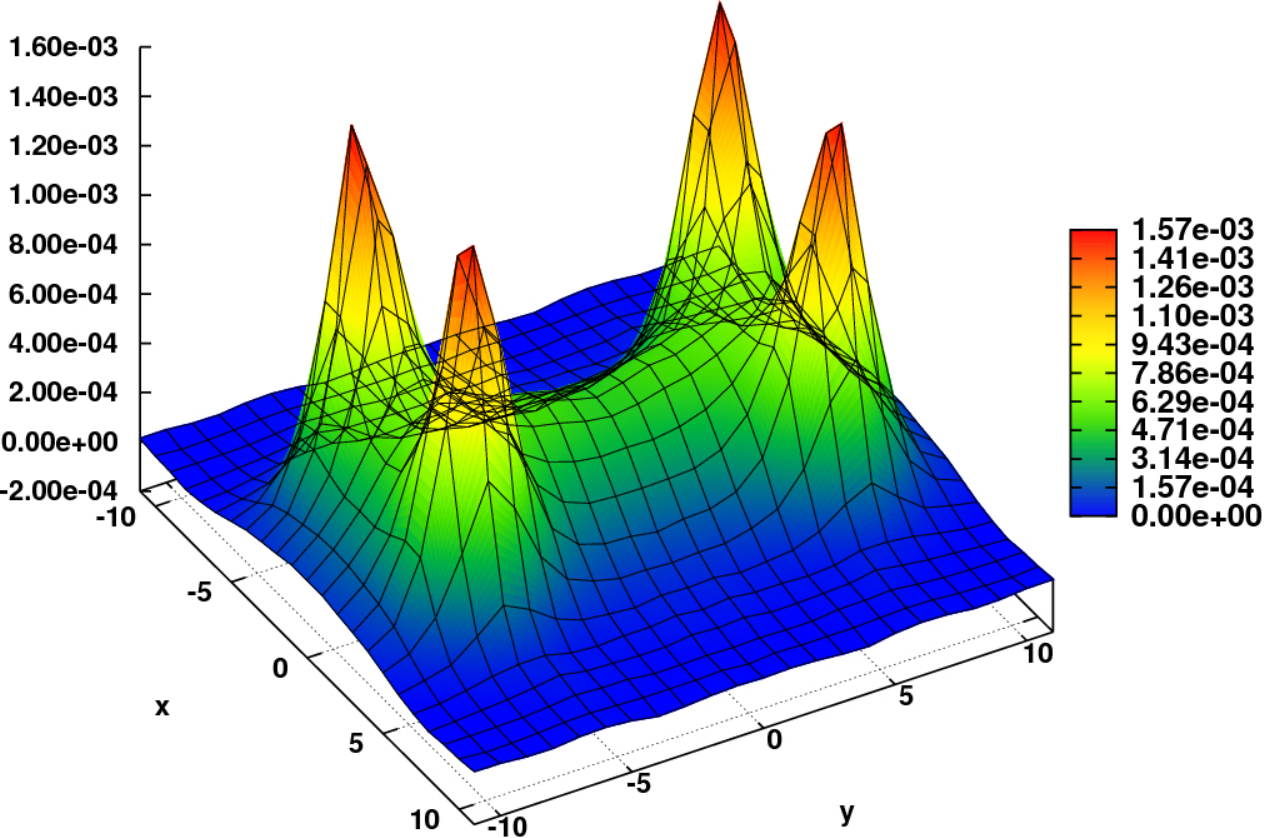}
\par\end{centering}}\par\end{centering}
    \caption{ We compare (a) the tetraquark flux tube (or string) model, the elementary flux tubes meet in two Fermat points, at an angle of  $\alpha=120^{\circ}$  to form a double-Y flux tube (except when this is impossible and the flux tube is X-shaped) with  (b) Lagrangian density 3D plot for $r_1=8, \ r_2=14$,  presented in lattice spacing units (colour online).}
    \label{fig:tq}
\end{figure}

On the theoretical side, the first efforts have been to search for bound states below the strong decay threshold
\cite{Beinker:1995qe,Zouzou:1986qh,Gelman:2002wf,Vijande:2007ix},
as it is apparent that the absence of a potential barrier may produce a large decay width to any open channel.
Recent investigations found that the presence of an angular momentum  centrifugal barrier may increase the stability of the system \cite{Karliner:2003dt,Bicudo:2010mv}.

In the last years, the static tetraquark potential has been studied in Lattice QCD computations \cite{Alexandrou:2004ak,Okiharu:2004ve,Bornyakov:2005kn}.
The authors concluded that when the quark-quark are well separated from the antiquark-antiquark,
the tetraquark potential is consistent with One Gluon Exchange Coulomb potentials plus a four-body confining potential,
suggesting the formation of a double-Y flux tube, as in Fig. \ref{fig:tq},
composed of five linear fundamental flux tubes meeting in two Fermat points \cite{Vijande:2007ix,Bicudo:2008yr,Richard:2009jv}.  A Fermat, or Steiner, point is defined as a junction minimizing the total length of strings, where linear individual strings join at $120^{\circ}$ angles.
When a quark approaches an antiquark, the minimum potential changes to a
sum of two quark-antiquark potentials, which indicates a two meson state.
This is consistent with the triple flip-flop potential, minimizing the length,
with either tetraquark flux tubes or meson-meson flux tubes,
of thin flux tubes connecting the different quarks or antiquarks
\cite{Vijande:2007ix,Bicudo:2010mv}.

Here we study the colour fields for the static tetraquark system
\cite{Cardoso:2010cf},
with the aim of observing
the tetraquark flux tubes suggested by these static potential computations.
The study of the colour fields in a tetraquark is important to discriminate between different
multi-quark Hamiltonian models.
Unlike the colour fields of simpler few-body systems, say mesons, baryons and hybrids,
\cite{Ichie:2002dy,Okiharu:2004tg,Cardoso:2009kz,Cardoso:2010kw},
the tetraquark fields have not been previously studied in lattice QCD.

\section{Computing Fields with the Wilson loop and the Plaquette}

\begin{figure}[t!]
\begin{center}
    \includegraphics[width=8cm]{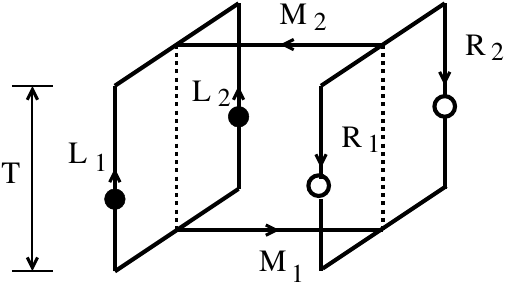}
\end{center}
\caption{Tetraquark Wilson loop as defined by Alexandrou et al
, and by Okiharu et al 
}
\label{fig:WL_TQ}
\end{figure}

To impose a static tetraquark, we utilize the respective Wilson loop
\cite{Alexandrou:2004ak,Okiharu:2004ve}
of Fig. \ref{fig:WL_TQ},
given by
$W_{4Q} = \frac{1}{3} \Tr \left( M_1 R_{12} M_2 L_{12} \right)$, where
\bea
R_{12}^{aa'} &=& \frac{1}{2}\epsilon^{abc}\epsilon^{a'b'c'}R_1^{bb'}R_2^{cc'}\ ,
\non \\
L_{12}^{aa'} &=& \frac{1}{2}\epsilon^{abc}\epsilon^{a'b'c'}L_1^{bb'}L_2^{cc'} \, .
\label{Wilson path}
\eea

The chromoelectric and chromomagnetic fields on the lattice are given by the
Wilson loop and plaquette expectation values,
\bea
\label{fields}
   \Braket{E^2_i(\mbf r)} &=& \Braket{P(\mbf r)_{0i}}-\frac{\Braket{W(r_1,r_2,T) \,P(\mbf r)_{0i}}}{\Braket{W(r_1,r_2,T)}}
 \\ \non
    \Braket{B^2_i(\mbf r)} &=& \frac{\Braket{W(r_1,r_2,T)\,P(\mbf r)_{jk}}}{\Braket{W}(r_1,r_2,T)}-\Braket{P(\mbf r)_{jk}} \, ,
\eea
where the $jk$ indices of the plaquette complement the index $i$ of the magnetic field,
and where the plaquette at position $\mbf r=(x,y,z)$ is computed at $t=T/2$,
\be
P_{\mu\nu}\left(\mbf r \right)=1 - \frac{1}{3} \ReC\,\Tr\left[ U_{\mu}(\mbf r) U_{\nu}(\mbf r+\mu) U_{\mu}^\dagger(\mbf r+\nu) U_{\nu}^\dagger(\mbf r) \right]\ .
\ee
The energy ($\mathcal{H}$) and lagrangian ($\mathcal{L}$) densities are then
computed from the fields,
\bea
   \Braket{ \mathcal{H}(\mbf r) } &=& \frac{1}{2}\left( \Braket{\mbf E^2(\mbf r)} + \Braket{\mbf B^2(\mbf r)}\right)\ , \\
    \label{energy_density}
   \Braket{ \mathcal{L}(\mbf r) } &=& \frac{1}{2}\left( \Braket{\mbf E^2(\mbf r)} - \Braket{\mbf B^2(\mbf r)}\right)\ .
    \label{lagrangian_density}
\eea

To produce the results presented in this work , we use 1121 quenched configurations in a $24^3 \times 48$ lattice at $\beta = 6.2$.
We also test that these configurations are already close to the continuum limit in a larger, $32^3\times 64$ lattice.
We present our results in lattice spacing units of $a$, with $a=0.07261(85)$ fm or $a^{-1}=2718\,\pm\, 32$ MeV.
We generate our configurations in NVIDIA GPUs of the FERMI series (480, 580 and Tesla 2070) with a SU(3) CUDA code
upgraded from our SU(2) combination of Cabibbo-Marinari
pseudoheatbath and over-relaxation algorithm \cite{Cardoso:2010di,ptqcd}.

\begin{figure}[t!]
\begin{center}
    \includegraphics[width=7.cm]{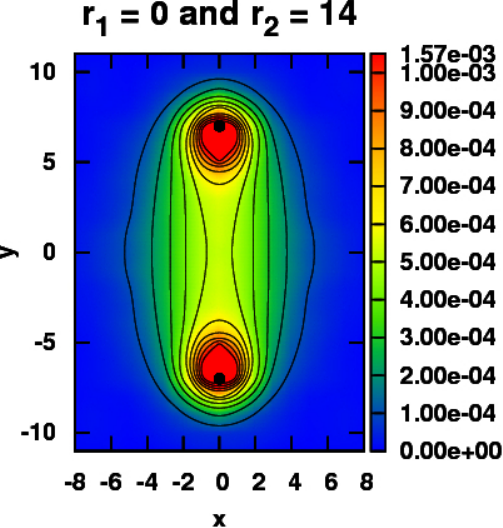}
    \includegraphics[width=7.cm]{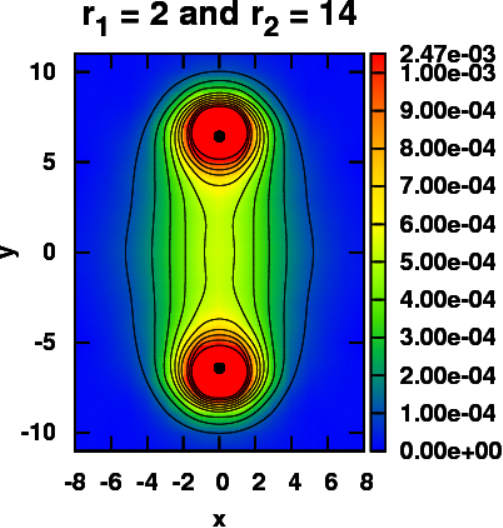}
\\
    \includegraphics[width=7.cm]{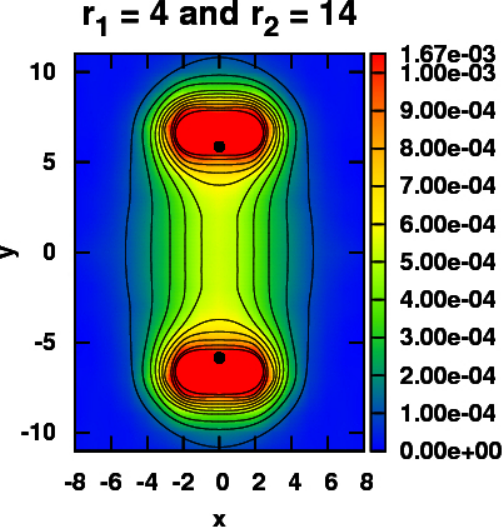}
    \includegraphics[width=7.cm]{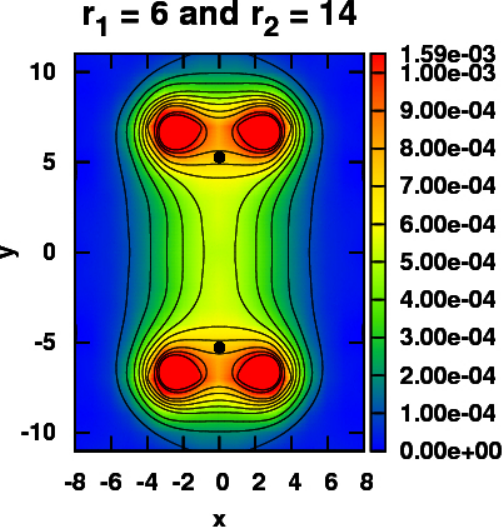}
    \caption{Lagrangian density for $r_2=14$ and $r_1$ from 0 to 6. The black dot points correspond to the Fermat points.
The results are presented in lattice spacing units (colour online).}
    \label{Act_ape_hyp_r2_14}
\end{center}
\end{figure}

To compute the static field expectation value, we plot the expectation value
    $ \Braket{E^2_i(\mbf r)} $ or   $\Braket{B^2_i(\mbf r)}$ as a function of the temporal extent $T$ of
    the Wilson loop.
In order to improve the signal to noise ratio of the Wilson loop, we use 50 iterations of APE Smearing
with $w = 0.2$ (as in
\cite{Cardoso:2009kz})
in the spatial directions and one iteration of hypercubic blocking (HYP) in the temporal direction.
 \cite{Hasenfratz:2001hp},
with $\alpha_1 = 0.75$, $\alpha_2 = 0.6$ and $\alpha_3 = 0.3$.
At sufficiently large $T$, the groundstate corresponding to the
studied quantum numbers dominates, and the expectation value of the fields tends to a horizontal plateau.
for each point $\mbf r$ determined by the plaquette position.
For the distances $r_1$ and $r_2$ considered, we find in the range of $T\in [3,12]$ in lattice units,
horizontal plateaux with a $\chi^2$ /dof $\in [0.3,2.0] $.
We finally compute the error bars of the fields with the jackknife method.

\section{ The Tetraquark fields }

\begin{figure}[t!]
\begin{centering}
    \subfloat[$\Braket{E^2}$\label{fig:TQ_EB_ape_hyp_r1_8_r2_14_E_Sim}]{
\begin{centering}
    \includegraphics[width=7.cm]{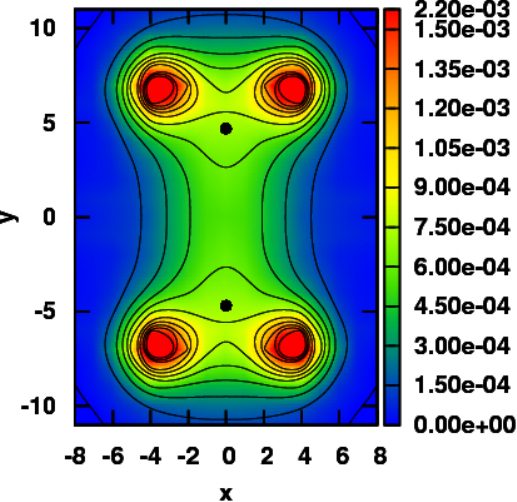}
\par\end{centering}}
    \subfloat[$-\Braket{B^2}$\label{fig:TQ_EB_ape_hyp_r1_8_r2_14_B_Sim}]{
\begin{centering}
    \includegraphics[width=7.cm]{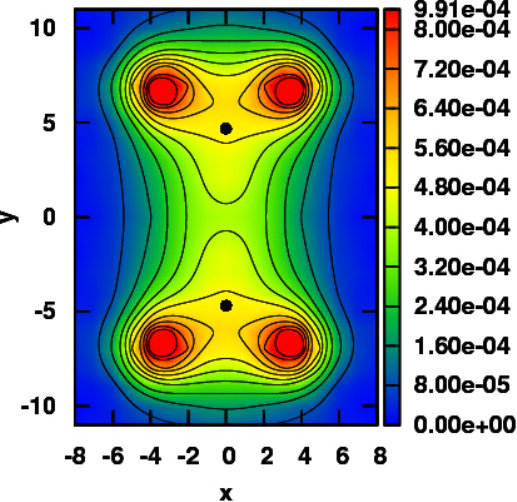}
\par\end{centering}}
\\
    \subfloat[Energy Density\label{fig:TQ_EB_ape_hyp_r1_8_r2_14_Energ_Sim}]{
\begin{centering}
    \includegraphics[width=7.cm]{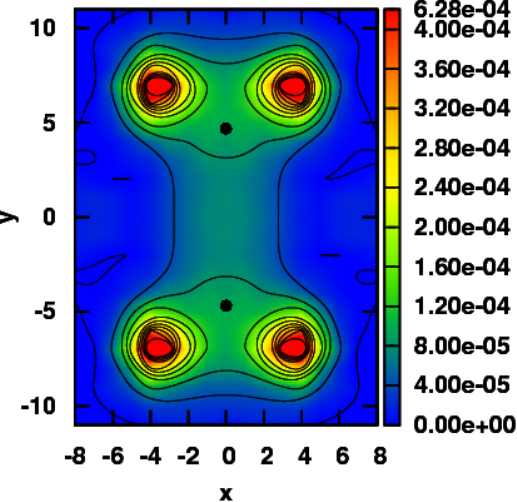}
\par\end{centering}}
    \subfloat[Lagrangian Density\label{fig:TQ_EB_ape_hyp_r1_8_r2_14_Act_Sim}]{
\begin{centering}
    \includegraphics[width=7.cm]{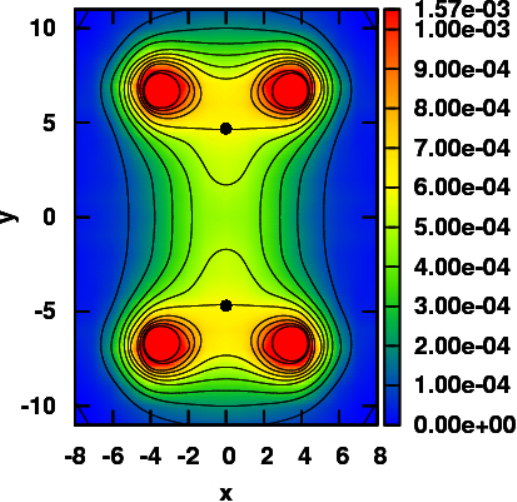}
\par\end{centering}}
\par\end{centering}
    \caption{Colour fields, energy density and Lagrangian density for $r_1=8$ and $r_2=14$. The black dot points correspond to the Fermat points.
The results are presented in lattice spacing units (colour online).}
    \label{ape_hyp_r1_8_r2_14}
\end{figure}

In our simulations, the quarks are fixed at $(\pm\,r_1/2,-r2/2,0)$ and the antiquarks at $(\pm\,r_1/2,r_2/2,0)$, with $r_1$ extending up to 8 lattice spacing units and $r_2$
extended up to 14 lattice spacing units, in order to include the relevant cases where $r_2 > \sqrt 3 r_1$.
Notice that in the string picture, at the line
 $r_2 = \sqrt 3 r_1$  in our $(r_1, \, r_2)$ parameter space, the transition between the
double-Y, or butterfly, tetraquark geometry in Fig. \ref{fig:tq1} to the meson-meson geometry  should occur.
The results are presented only for the $xy$ plane since the quarks are in this plane and the results with $z\neq 0$ are less interesting for this study.
The flux tube fields can be seen in Fig. \ref{fig:TQ_EB_ape_hyp_r1_8_r2_14_Act_3D_Sim}, \ref{Act_ape_hyp_r2_14} and \ref{ape_hyp_r1_8_r2_14}.
Theses figures exhibit clearly tetraquark double-Y, or butterfly, shaped flux tubes. The
flux tubes have a finite width, and are not infinitely thin as in the string models inspiring the Fermat points and the triple flip-flop potential, but nevertheless the junctions are close to the Fermat points, thus justifying the use of string models for the quark confinement in constituent quark models.

\begin{figure}[t!]
\begin{center}
    \includegraphics[width=6.cm]{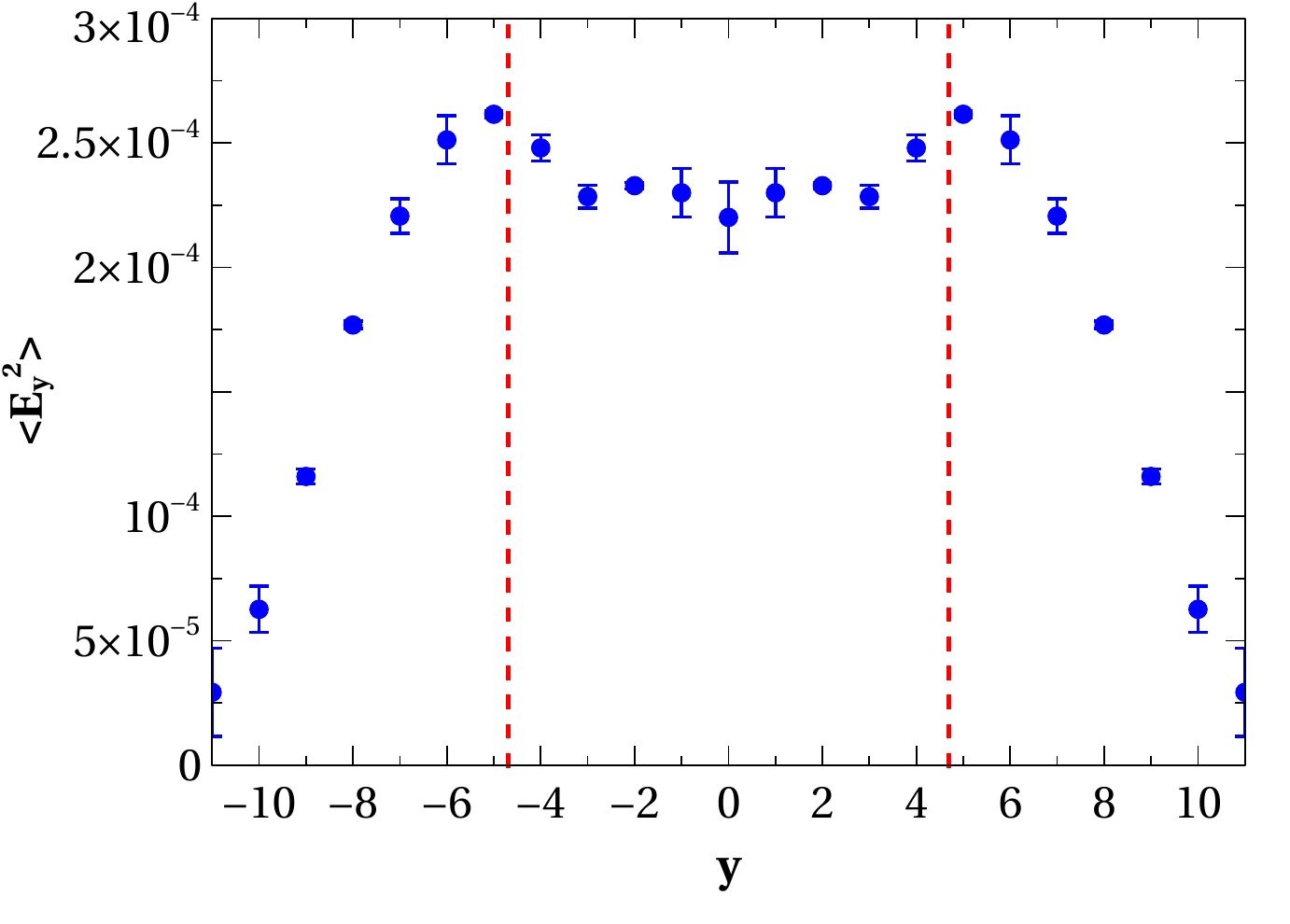}
    \includegraphics[width=6.cm]{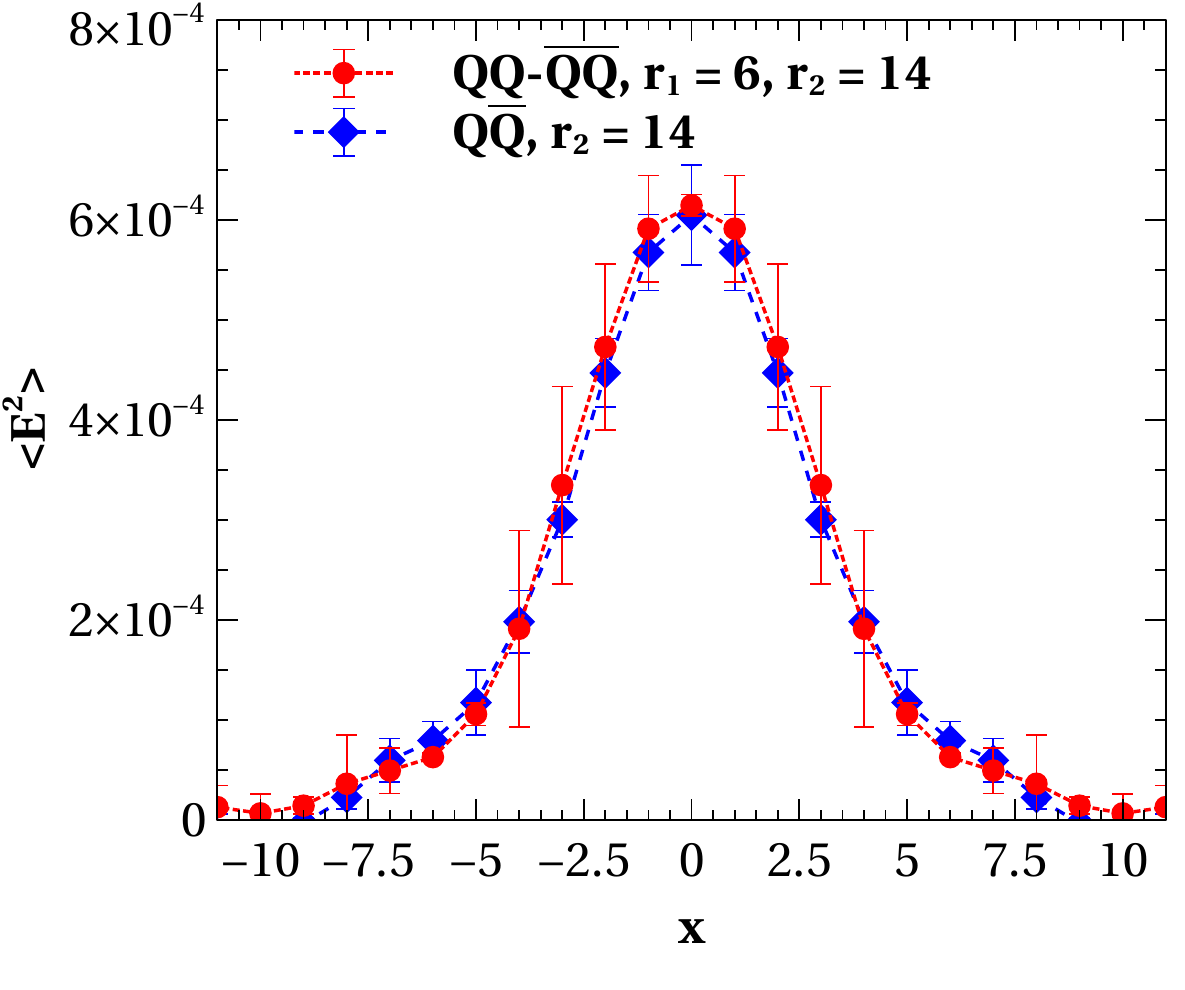}
  \caption{(a) $\Braket{E_y^2}$ in the central axis  $x=0$ for $r_1=8, \ r_2=14$. We show with vertical dashed lines the location of the two Fermat points. 
(b) Profile cut at $y=0$ of the chromoelectric field for the tetraquark and quark-antiquark systems in the middle of the flux tube. The results are presented in lattice spacing units (colour online).}
    \label{fig:TQ_QQ_profile}
\end{center}
\end{figure}

In Fig. \ref{fig:TQ_QQ_profile} (a) , we plot the chromoelectric field along the central flux tube, $\Braket{E_y^2}$ at $x=0$, for $r_1=8, \, r_2=14$. As expected, the chromoelectric field along $y$ is in agreement with the position of the Fermat points. The chromoelectric field along the $x=0$ central axis is maximal close to the Fermat points situated
at $x \simeq -4.69$ and at $x \simeq 4.69$, flattens in the middle of the flux tube.
Outside the flux tube,  the chromoelectric field is almost residual.

In Fig. \ref{fig:TQ_QQ_profile} (b), we compare the chromoelectric field for the tetraquark and the quark-antiquark system in the middle of the flux tube between the (di)quark and the (di)antiquark.
As can be seen, for our larger distance $r_2=14$ where the source effects are small,
the chromoelectric field is identical up to the error bars, and this confirms that the tetraquark 
flux tube is composed of  a set of fundamental flux tubes with Fermat junctions.

To check which of the colour structures, tetraquark or meson-meson, produces the groundstate flux tube, we study the $\chi^2 /$dof of the $T$ plateaux. 
Surprisingly, event at distances as small as $r_2 \simeq {1 \over 2}  r_1 \sqrt 3$,  where the flip-flop potential favours the two-meson flux tube, we find $T$ plateaux with a good   $\chi^2$ /dof. This shows that the mixing between the tetraquark flux tube and the meson-meson flux tube is small, and it is possible to study clear tetraquark flux tubes even at relatively small quark-antiquark distances.

\section{Foreword}

\begin{figure}[t!]
\begin{center}
\includegraphics[width=7.cm]{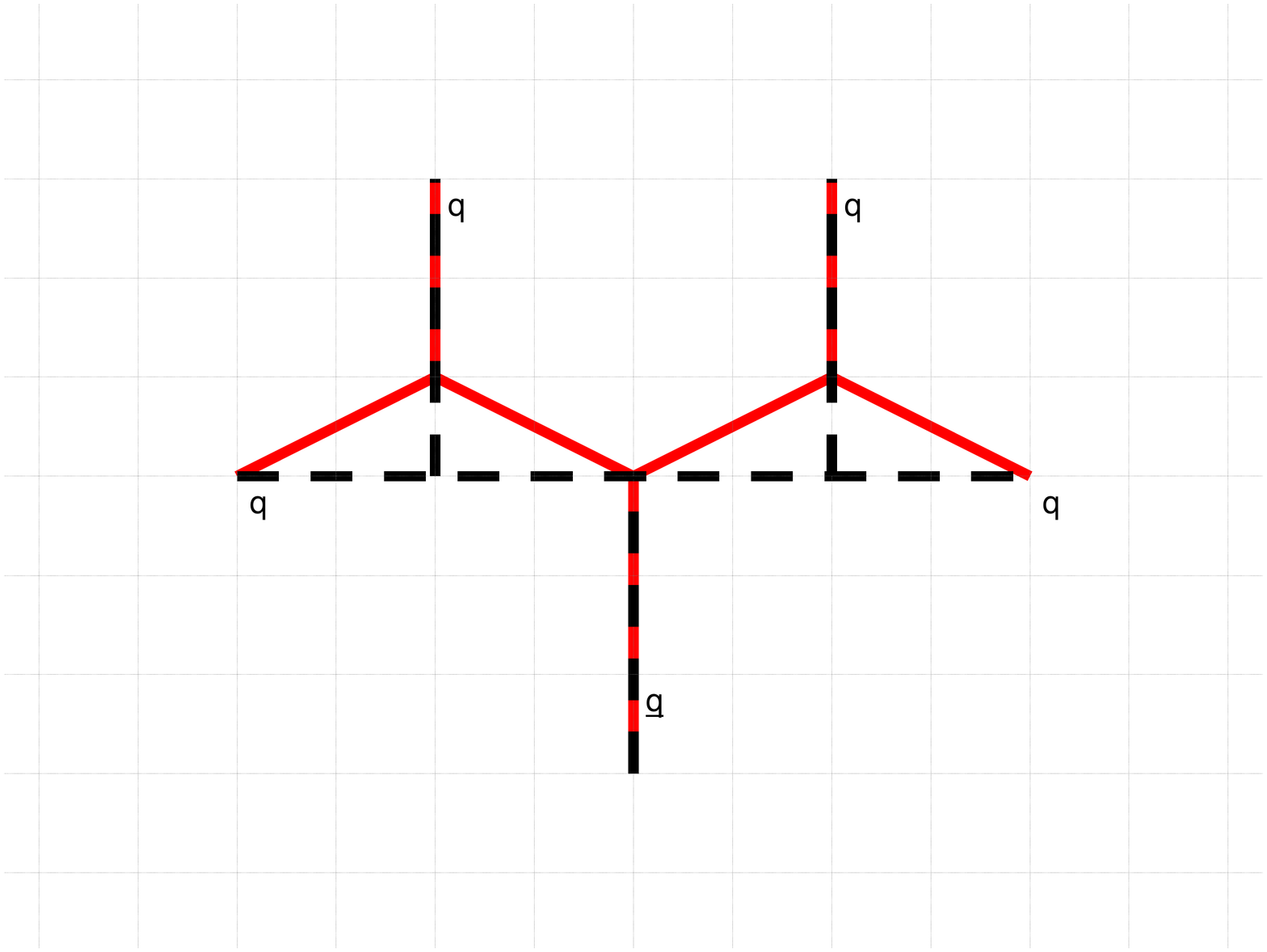}
\includegraphics[width=8.cm]{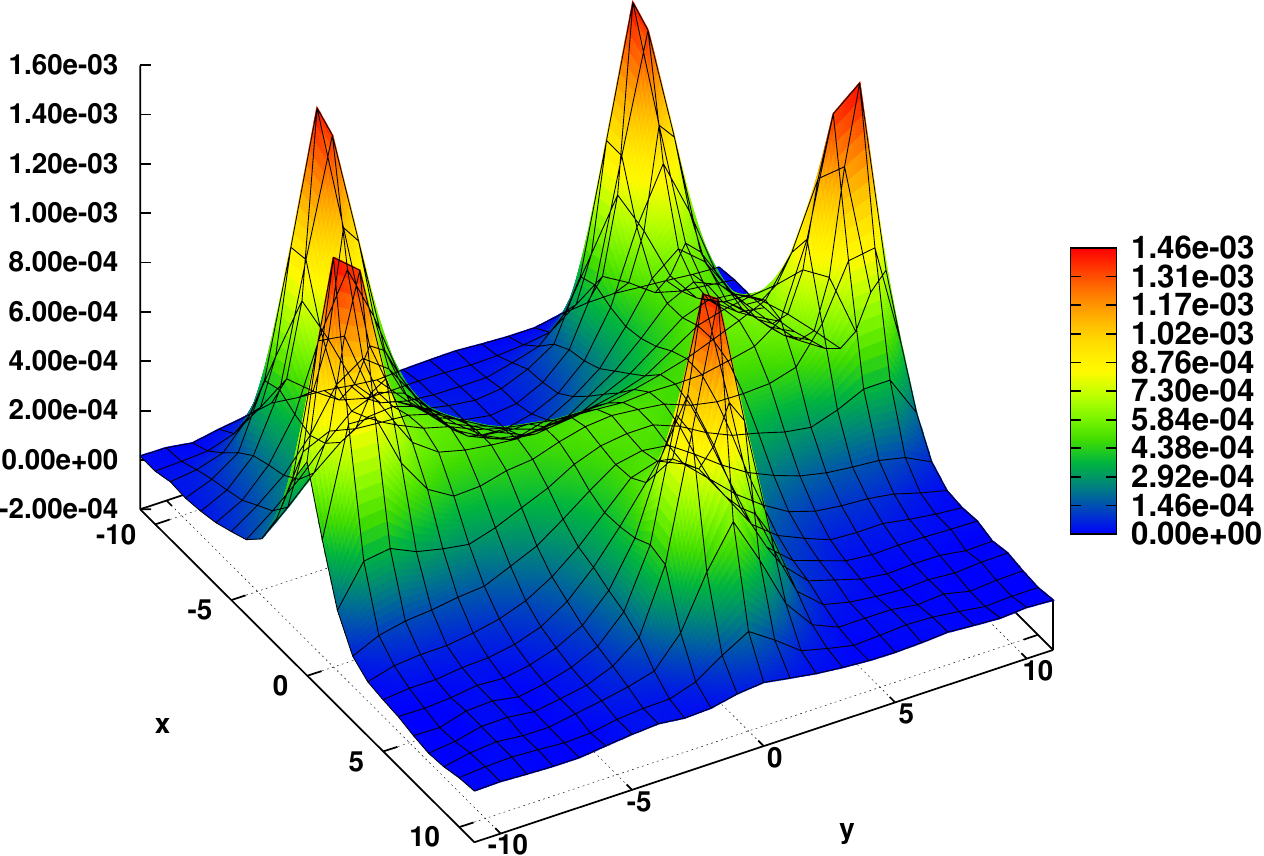}
\end{center}
    \caption{Wilson loop geometry, and preliminary result for the field, (Lagrangian density) for the static pentaquark.
The results are presented in lattice spacing units (colour online).}
    \label{geom5}
\end{figure}

\begin{itemize}
\item The flux tubes remain interesting in Lattice QCD, our results support the string model of confinement, in particular for the tetraquark static potential 
\item The mixing between the tetraquark and meson-meson flux-tubes is small, which may contribute for narrower tetraquark resonances.
\item We are now studying in more detail other flux tubes, as for the pentaquark in Fig. \ref{geom5}.
 \item We will soon have codes for SU(2), SU(3), SU(4), etc, available on our webpage \cite{Cardoso:2010di,ptqcd}.
\end{itemize}

\vskip 0.3in
{\it
This work was partly funded by the FCT contracts, PTDC/FIS/100968/2008,  CERN/FP/109327/2009, CERN/FP/116333/2010 and CERN/FP/116383/2010.

Nuno Cardoso is also supported by FCT under the contract SFRH/BD/44416/2008.
}

\bibliographystyle{ieeetr}
\bibliography{bib_proc}

\end{document}